\documentclass[DIV12,11pt]{scrartcl}

\usepackage{amsthm,amssymb,amsmath}
\usepackage[noblocks]{authblk}
\usepackage{graphics}
\usepackage{multirow}

\setkomafont{title}{\normalfont \LARGE \bfseries}
\setkomafont{section}{\normalfont \Large \bfseries \boldmath}
\setkomafont{subsection}{\normalfont \large \bfseries}
\setkomafont{subsubsection}{\normalfont \normalsize \bfseries}
\setkomafont{descriptionlabel}{\itshape}
\setkomafont{paragraph}{\normalfont \bfseries \boldmath}

\newtheorem{theorem}{Theorem}
\newtheorem{lemma}[theorem]{Lemma}

\title{Improved Space-efficient Linear Time Algorithms for Some Classical Graph Problems}
\author[1]{Sankardeep Chakraborty}
\author[2]{Seungbum Jo}
\author[3]{Srinivasa Rao Satti}
\affil[1]{The Institute of Mathematical Sciences, HBNI, Chennai, India. sankardeep@imsc.res.in}
\affil[2]{University of Siegen, Siegen, Germany. seungbum.jo@uni-siegen.de}
\affil[3]{Seoul National University, Seoul, South Korea. ssrao@cse.snu.ac.kr}

\date{\vspace*{-2em}}

\begin{document}

\maketitle

\thispagestyle{empty}

\begin{abstract}
We provide space-efficient linear time algorithms for computing bridges, topological sorting, and strongly connected components improving on several recent results of Elmasry et al. [STACS'15], Banerjee et al. [COCOON'16] and Chakraborty et al. [ISAAC'16]. En route, we also provide another DFS implementation with weaker input graph representation assumption without compromising on the time and space bounds of the earlier results of Banerjee et al. [COCOON'16] and Kammer et al. [MFCS'16].
\end{abstract}

%\section{My First Section}

\section{Introduction} \label{bicon}
Since the early days of designing graph algorithms, researchers have developed several approaches for testing whether a given undirected (or directed) graph $G=(V,E)$ with $n$ vertices and $m$ edges is (strongly connected) biconnected and/or $2$-edge connected, and finding cut vertices and/or bridges of $G$. All of these methods use depth-first search (DFS) as the backbone to design the main algorithm. The classical linear time algorithms due to Tarjan~\cite{Tarjan72,Tarjan74} computes the so-called ``low-point" values (which are defined in terms of a DFS-tree of $G$) for every vertex $v$, and checks some conditions using that to determine whether $G$ has the desired property. 
%The concept of ``low-point" was introduced first by Tarjan in~\cite{Tarjan72} and since then it has been applied for designing many other efficient graph algorithms.
%Since then 
%the algorithm of Tarjan~\cite{Tarjan72}, 
There are other linear time algorithms as well for these problems (see~\cite{Schmidt13} and all the references therein).
%also have been designed 
%for all the problems mentioned before 
%(e.g.,~\cite{Ebert83,EvenT76} etc). 
%In particular, Brandes~\cite{Brandes02} and Gabow~\cite{Gabow00} gave much simpler algorithms for testing biconnectivity and/or strong connectivity along with computing biconnected and/or strongly connected components by using simple path-generating rules instead of low-points.
%; they call these algorithms path-based. 
%Another algorithm, due to Schmidt~\cite{Schmidt13}, is based on chain decomposition of graphs to determine biconnectivity and $2$-edge connectivity. 
All of these classical algorithms take $O(m+n)$ time and $O(n)$ words (our model of computation is the standard word RAM model with word size $w = \Omega (\lg n)$ bits) of space. Our aim is to improve the space bounds of these algorithms without increasing the running time.

\subsection{Motivation and Related Work}
%{\bf Motivation and Related Work:} 
Motivated mainly by the ``big data" phenomenon among others, recently there has been a surge of interest in improving the space complexity of the fundamental linear time graph algorithms by paying little or no penalty in the running time i.e., reducing the working space of the classical graph algorithms (which generally take $O(n \lg n)$ bits) to $o(n \lg n)$ bits without compromising on time.
%or slightly more with little or no penalty in running time. 
Towards this, Elmasry et al.~\cite{ElmasryHK15} gave, among others, an implementation for DFS taking $O(m+n)$ time and $O(n \lg \lg n)$ bits of space.
%, improving the result of Asano et al.~\cite{AsanoIKKOOSTU14}. 
For sparse graphs (when $m=O(n)$), the space bound was improved further to $O(n)$ bits keeping the same linear time in~\cite{BanerjeeC016}. Banerjee et al.~\cite{BanerjeeC016} gave, among others, a space efficient implementation for performing BFS using just $2n+o(n)$ bits of space and linear time, improving upon the result of~\cite{ElmasryHK15}. Such
%Space efficient 
algorithms for a few other graph problems also have been considered recently~\cite{BanerjeeCRRS2015,Chakraborty0S16,ChakrabortyS17,CRS17,KammerKL16}. 

\subsection{Our Results}
%We assume the standard word RAM model with word size $w = \Omega (\lg n)$ bits, and 
We assume that the input graph $G$, which is represented using \textit{adjacency array}~\cite{BanerjeeC016,Chakraborty0S16,ElmasryHK15,KammerKL16}, i.e., $G$ is represented by an array of length $|V|$ where the $i$-th entry stores a pointer to an array that stores all the neighbors of the $i$-th vertex,
%as the standard adjacency list, 
is given in a read-only memory with a limited read-write working memory, and write-only output. We count space in terms of the number of bits in workspace used by the algorithms. 
%We assume that the input graphs $G$ are represented using \textit{adjacency array}~\cite{BanerjeeC016,ElmasryHK15,Chakraborty0S16,KammerKL16}, i.e., $G$ is represented by an array of length $|V|$ where the $i$-th entry stores a pointer to an array that stores all the neighbors of the $i$-th vertex. 
Our main goal here is to improve the space bounds of some of the classical and fundamental graph algorithms. We summarize all our main results in Table $1$.
%~\ref{table}. 
In this paper, basically we complete the full spectrum of results regarding the 
space bounds for these problems keeping the running time linear by 
providing/improving the missing/existing algorithms in the recent space 
efficient graph algorithm literature. Due to lack of space, we provide only sketches of our proofs.

\begin{table}
\label{table}
\centering
\scalebox{0.8}{
\begin{tabular}{|c | c | c | c |c |c |c|}
\hline
\multirow{2}{*}{Time} & \multirow{2}{*}{Space (in bits)} & \multirow{2}{*}{DFS} 
&  Testing biconnectivity & Testing 2-edge 
connectivity &Topological & Testing strong\\
& & & \& reporting cut vertices & \& reporting bridges  & sort & 
connectivity\\ 
\hline
$O(n+m)$& $O(n\lg{n})$ & \cite{CLRS}& \cite{Tarjan72} &\cite{Tarjan74} &\cite{CLRS} & \cite{Tarjan72}\\
$O(n+m)$& $O(n+m)$ &\cite{BanerjeeC016,KammerKL16} & \cite{BanerjeeC016} & \cite{BanerjeeC016} & This paper  & This paper\\
$O(n+m)$& $O(n\lg(m/n))$ & \cite{Chakraborty0S16}&\cite{Chakraborty0S16} &\cite{Chakraborty0S16} & This paper & This paper\\
$O(n+m)$& $O(n\lg{\lg{n}})$ &\cite{ElmasryHK15} & \cite{KammerKL16}& This paper & \cite{ElmasryHK15}& \cite{ElmasryHK15}\\
%$O(m\lg^c{n}\lg{\lg{n}})$ & $O(n)$ & & & & & \\
\hline
\end{tabular}}
\caption{Summary of our results.}
\label{table:summary}
\end{table}

\section{Testing $2$-Edge Connectivity and Finding Bridges}
In an undirected graph $G$, a bridge is an edge that when removed (without removing the vertices) from a graph creates more components than previously in the graph. A (connected) graph with at least two vertices is $2$-edge-connected if and only if it has no bridge. Let $T$ denote the DFS tree of G. Following Kammer et al.~\cite{KammerKL16}, we call a tree edge $(u, v)$ of $T$ with $u$ being the parent of $v$ \textit{full marked} if there is a back edge from a descendant of $v$ to a strict ancestor of $u$, \textit{half marked} if it is not full marked and there exists a back edge from a descendant of $v$ to $u$, and \textit{unmarked}, otherwise. They use this definition to 
%Kammer et al.~\cite{KammerKL16} recently 
prove the following: 
%characterization: 
(i) every vertex $u$ (except the root $r$) is a cut vertex exactly if at least one of the edges from $u$ to one of its children is either an unmarked edge or a half marked edge, and (ii) root $r$ is a cut vertex exactly if it has at least two children in $T$. Based on the above characterization, they gave $O(m+n)$ time and $O(n \lg \lg n)$ bits algorithm to test/report if $G$ has any cut vertex. Our main observation is that we can give a similar characterization for bridges in $G$, and essentially using a similar implementation, we can also obtain $O(m+n)$ time and $O(n \lg \lg n)$ bits algorithms for testing $2$-edge connectivity and reporting bridges of $G$. We start with the following lemma.
%Towards that, we first note the following, 
\begin{lemma}\label{edge}
A tree edge $e=(u,v)$ in $T$ is a bridge of $G$ if and only if it is unmarked.
\end{lemma}
\noindent
\textbf{Proof sketch:} If $e$ is unmarked, then no descendants of $v$ reaches $u$ or any strict ancestor of $u$, so deleting $e$ would result in disconnected graph, thus $e$ has to be a bridge. On the other direction, it is easy to see that if $e$ is a bridge, it has to be an unmarked edge. \qed

Now we state our theorem below.
\begin{theorem}\label{edge_main}
Given an undirected graph $G$, in $O(m+n)$ time and $O(n \lg \lg n)$ bits of space we can determine whether $G$ is $2$-edge connected. If $G$ is not $2$-edge connected, then in the same amount of time and space, we can compute and output all the bridges of $G$.
\end{theorem}

\noindent
\textbf{Proof sketch:} Using Lemma~\ref{edge} and the similar implementation of  using stack compression and other tools of the algorithm provided in Section $3.2$ of Kammer et al.~\cite{KammerKL16} with few modifications, we can prove 
%Theorem~\ref{edge_main}. 
the theorem.\qed

Note that the space bound of Theorem~\ref{edge_main} improves the results of~\cite{BanerjeeC016} and~\cite{Chakraborty0S16} for sufficiently dense graphs (when $m=\omega(n \lg \lg n)$ and $m=\omega(n \lg^{O(1)}n)$ respectively) while keeping the same linear runtime (see Table~$1$).

\section{DFS without Cross Pointers}
Banerjee et al.~\cite{BanerjeeC016} and subsequently Kammer et al.~\cite{KammerKL16} gave $O(m+n)$ bits and $O(m+n)$ time implementations of DFS improving on the bounds of~\cite{ElmasryHK15} for sparse graphs. But both of these DFS implementations assume that the input graph is represented using the \textit{adjacency array} along with \textit{cross pointers} i.e., for undirected graphs, every neighbour $v$ in the adjacency array of a vertex $u$ stores a pointer to the position of vertex $u$ in the adjacency array of $v$. See~\cite{ElmasryHK15} for detailed definitions for directed graphs. We emphasize that  this input assumption can double the space usage, compared to the raw adjacency array in worst case. In what follows, we provide the proof sketch of a DFS implementation taking the same time and space bounds as that of~\cite{BanerjeeC016,KammerKL16} but without using the cross pointers. Our main theorem is as follows.

\begin{theorem}\label{modi_dfs}
%A DFS traversal of $G$ can be performed using $O(m+n)$ bits and $O(m+n)$ time.
Given a directed or undirected graph $G$, represented as adjacency array, we can perform DFS traversal of $G$ using $O(m+n)$ bits and $O(m+n)$ time.
\end{theorem}

\noindent
\textbf{Proof sketch:} We essentially modify the proof of~\cite{BanerjeeC016} 
which uses a bitvector $A$ of length $O(m+n)$ having one to one mapping with the unary encoding of the degree sequence to mark the tree edges, and subsequently uses cross 
pointers to find the parent of any vertex during backtracking as well as 
starting with next unvisited vertex after backtracking. We note that we can represent the parents of all the vertices in another bitvector $P$ of length $O(m+n)$ (parallel to $A$). Now to perform backtracking efficiently, we could 
use the constant time \textit{append only} structure (also with constant time 
rank/select) of Grossi et al.~\cite{GrossiO12} along with the $P$ array. With 
these modifications, we could get rid of cross pointers without compromising on 
the running time and space bound of the earlier algorithms.\qed

\section{Testing Strong Connectivity and Topological Sorting}
Towards giving improved space efficient algorithms for strong connectivity (SC) and topological sorting (TS), we first improve Lemma $4.1$ of~\cite{ElmasryHK15} which says the following: if DFS of a directed graph $G$ takes $T(n,m)$ time and $S(n,m)$ space, then we can output the vertices of $G$ in reverse postorder of the DFS tree $T$ of $G$ taking $O(T(n,m))$ time and $O(S(n,m)+n \lg \lg n)$ space. Combining this lemma with the classical algorithms for SC and TS~\cite{CLRS} 
%whose main procedure steps through the vertices of $G$ in reverse postorder, 
they obtained $O(n \lg \lg n)$ bits and $O(m+n)$ time algorithms for both these problems. We improve these by showing the following,
\begin{theorem}\label{worse}
If DFS of a directed graph $G$ takes $T(n,m)$ time and $S(n,m)$ space, then the vertices of $G$ can be output in reverse postorder with respect to a DFS forest of $G$ taking $O(T(n,m))$ time and $O(S(n,m)+ m+n)$ space. As a result, we can also solve SC and TS in $O(m+n)$ time using $O(n+m)$ bits of space.
\end{theorem}
\noindent
\textbf{Proof sketch:} We use the DFS algorithm of Theorem~\ref{modi_dfs} to first mark all the tree edges in the array $A$. Now we start with the rightmost leaf vertex of the DFS tree and use rank/select operations~\cite{GrossiO12} on $A$ and $P$ (as defined in the proof of Theorem~\ref{modi_dfs}) carefully to traverse the tree in reverse direction (along with standard DFS backtracking etc) to generate reverse postorder sequence. Now using this as the back bone of the classical algorithms, we obtain $O(m+n)$ bit and $O(n+m)$ time algorithms for SC and TS.\qed

%\begin{corollary}\label{worse_tmn}
%We can solve SC and TS using $O(m+n)$ time and $O(n+m)$ bits of space.
%\end{corollary}

%Now using Lemma~\ref{worse} as the back bone of the classical algorithms, we obtain $O(m+n)$ time and $O(n+m)$ bits algorithm for SC and TS. 
Theorem~\ref{worse} improves the result of~\cite{ElmasryHK15} for sparse (when 
$m=O(n)$) graphs. Now if we use the DFS algorithm of Chakraborty et 
al.~\cite{Chakraborty0S16} and modify it suitably to perform the traversal of 
the DFS tree in reverse, we obtain the following result.
%If we combined the data structure of Chakraborty et al., we can obtain a following theorem (and remove proof) This also sketches the proof of the entries in the second row of the last two columns of Table $1$. To prove the entries the in the third row and last two columns, we show the following. %We omit the details of the proof.
\begin{theorem}\label{worse1}
If DFS of a directed graph $G$ takes $T(n,m)$ time and $S(n,m)$ space, then the vertices of $G$ can be output in reverse postorder with respect to a DFS forest of $G$ taking $O(T(n,m))$ time and $O(S(n,m)+ n\lg(m/n))$ space. As a result, we can also solve SC and TS using $O(m+n)$ time and $O(n\lg(m/n))$ bits.
\end{theorem}

\end{document}